\documentclass[]{emulateapj}

\usepackage{amsmath}
\usepackage{epsfig}
\usepackage{url}

\newfont{\myfont}{cmmib10}

\newfont{\myfontsmall}{cmmib8}

\DeclareSymbolFont{cmmi}{OML}{cmm}{m}{it}
\DeclareMathSymbol{v}{\mathalpha}{cmmi}{"76}

\shorttitle{Detecting MSPs at the Galactic Center}
\shortauthors{}

\begin{document}

\title{On Detecting Millisecond Pulsars at the Galactic Center}

\author{Jean-Pierre Macquart\altaffilmark{1,2} and Nissim Kanekar\altaffilmark{3}}

\altaffiltext{1}{ICRAR/Curtin University, Curtin Institute of Radio Astronomy, Perth WA 6845, 
Australia {\it J.Macquart@curtin.edu.au}}
\altaffiltext{2}{ARC Centre of Excellence for All-Sky Astrophysics (CAASTRO)}
\altaffiltext{3}{Swarnajayanti Fellow, National Centre for Radio Astrophysics, 
Tata Institute of Fundamental Research, Ganeshkhind, Pune - 411007, India}
\begin{abstract}
The lack of detected pulsars at the Galactic Center (GC) region is a long-standing mystery. We 
argue that the high stellar density in the central parsec around the GC is likely to result 
in a pulsar population dominated by millisecond pulsars (MSPs), similar to the situation in 
globular cluster environments. Earlier GC pulsar searches have been largely insensitive to 
such an MSP population, accounting for the lack of pulsar detections.  We estimate the best
search frequency for such an MSP population with present and upcoming broad-band radio telescopes 
for two possible scattering scenarios, the ``weak-scattering'' case suggested by the 
recent detection of a magnetar close to the GC, and the ``strong-scattering'' case, with the 
scattering screen located close to the GC. The optimal search frequencies are $\approx 8$~GHz
(weak-scattering) and $\approx 25$~GHz (strong-scattering), for pulsars with periods $1-20$~ms, 
assuming that GC pulsars have a luminosity distribution similar to that those in the rest of 
the Milky Way. We find that $10-30$~hour integrations with the Very Large Array and the Green Bank 
Telescope would be sufficient to detect MSPs at the GC distance in the weak-scattering case. However,
if the strong-scattering case is indeed applicable to the GC, observations with the full Square 
Kilometre Array would be needed to detect the putative MSP population.
\end{abstract}


\section{Introduction}
The allure of testing General Relativity near a supermassive black hole and investigating 
the latter's accretion environment \citep[see, e.g.,][]{pfahl04,liu12} has provided the motivation 
for numerous searches for radio pulsars at the Galactic Center (GC) over the last two decades 
\citep[e.g.][]{johnston06,deneva10,macquart10}.  Yet, despite predictions that $100 - 1000$ 
radio pulsars with orbital periods $\lesssim 100$~years are orbiting Sgr\,A* 
\citep[][]{pfahl04,wharton12,chennamangalam14,zhang14}, not a single normal pulsar has so far 
been detected within $10^\prime$ of  Sgr A*.

A long-standing problem related to these pulsar surveys relates to the scattering environment 
at the GC.  The extreme angular broadening exhibited by Sgr\,A* and OH/IR stars
in the GC region \citep[e.g.][]{lo85,frail94}, coupled with inferences about the 
distribution of the scattering material, led to the conclusion that the pulse smearing caused 
by turbulence along the line of sight to the GC is extreme, with a temporal 
smearing time of $\approx 10^3\, \nu_{\rm GHz}^{-4}$\,s, where $\nu_{\rm GHz}$ is the 
observing frequency in GHz \citep[][]{cordes97,lazio98}. Such a large temporal smearing time would obstruct the 
detection of pulsed radio emission at the usual low frequencies at which pulsar searches 
are usually carried out (due to their steep radio spectrum). Indeed, for normal pulsars, with 
periods $\approx 0.5$~s, the temporal smearing would become negligible only at frequencies 
above $15$\,GHz \citep[][]{macquart10}, where the steepness of the pulsar emission renders 
them weak and hence difficult to detect, even with today's most sensitive telescopes.

The recent detection of the GC magnetar PSR\,J1745-29, located within $3^{\prime \prime}$ 
of the position of Sgr\,A* \citep[][]{kennea13,mori13}, has caused an upheaval in the 
field, lending renewed impetus to searches for pulsars close to Sgr\,A*. \citet{spitler14} 
reported the detection of pulsed emission from the magnetar at frequencies 
as low as 1.1\,GHz, inferring a temporal smearing timescale of only 
$(1.3\pm 0.2) \nu_{\rm GHz}^{-4}$\,s. The effect of pulse smearing on at least one object 
in the GC environment is thus three orders of magnitude smaller than that 
inferred from earlier studies. The magnetar also exhibits the same degree of angular broadening 
as Sgr\,A* \citep[][]{bower14}, suggesting that it lies behind the same hyperturbulent 
scattering region. This hypothesis is supported by the high rotation measure of the magnetar 
\citep[][]{eatough13,shannon13}, which indicates that its proximity to Sgr\,A* on the 
sky is not merely a chance alignment.

The detection of the magnetar close to the GC suggests that the pulse smearing towards 
the GC is also relatively 
benign, assuming that the scattering screen is uniform. This would imply that radiation from 
ordinary pulsars at the GC should be readily detectable at frequencies $\gtrsim 3\,$GHz. 
However, if the effects of pulse smearing are similarly small over the entire GC region, 
previous surveys should have detected a significant fraction of the pulsar distribution 
with ``normal'' spin periods \citep[][]{wharton12,dexter14}. This dearth of pulsar 
detections is particularly acute given that magnetars are believed to represent 
only $\sim 0.2\%$ of all radio pulsars \citep[][]{olausen14,chennamangalam14}. The lack of 
``slow'' pulsar detections suggests that they either constitute an unusually small fraction 
of the entire GC pulsar population, or are significantly under-luminous relative to 
the population of slow pulsars that have been detected elsewhere in the Galaxy.  
  
In passing, we note that \citet{chennamangalam14} used a Bayesian analysis combined with 
an assumed log-normal pulsar luminosity function to find that existing pulsar surveys at 
the GC are not sufficiently deep to eliminate the possibility that a substantial population 
of non-recycled low-luminosity pulsars might exist at the GC.  While this is broadly correct, 
the shape of the pulsar luminosity function is not at all well-constrained at the low-luminosity 
end. As such, there is considerable error in the extrapolation of the distribution to low 
luminosities, and hence on such estimates of the total size of the normal pulsar population in 
the central parsec. For example, an uncertainly of only 20\% in the two parameters of the 
log-normal distribution \citep{bagchi11,chennamangalam14} yields an uncertainty of two orders of 
magnitude in the survey completeness. Similar estimates of the number of potential MSPs at 
the GC are plagued with an even greater degree of uncertainty because the luminosity function 
of recycled pulsars in globular clusters is even less well known. In the present analysis,
we will adopt a pragmatic approach to the pulsar luminosity distribution of pulsars: our detection 
arguments are made only with reference to the known luminosities of detected Galactic pulsars. 
  
In this paper, we explore the reasons for the paucity of pulsar detections at the Galactic 
Center. In \S\,\ref{sec:msp}, we discuss the nature of the pulsar population at the GC region,
and argue that it is likely to be dominated by recycled millisecond pulsars (MSPs);
present pulsar surveys of the GC have been largely insensitive to such a population. 
Next, we present in \S\,\ref{SNSection} detailed calculations of the expected signal-to-noise 
ratio for a wide-bandwidth search for MSPs 
at the Galactic Center with present and future telescopes, incorporating the frequency 
dependence of temporal smearing and sky temperature across the different observing bands. 
Finally, the results of this work are summarized in \S\,\ref{ConclusionsSection}.

\section{A Millisecond Pulsar Population at the Galactic Center} 
\label{sec:msp}

Over the last two decades, several searches have been carried out for pulsars at
the GC, mostly at frequencies between 1.4~GHz and 8~GHz 
\citep[e.g.][]{johnston95,johnston06,kramer00,deneva09,deneva10,bates11}. Based on the high expected 
temporal smearing, \citet{macquart10} argued that higher observing frequencies are more 
amenable to the discovery of pulsars, with the optimal frequency range for searches for 
``normal'' pulsars -- those with periods of $\approx 0.5$~seconds -- being $10 - 16$~GHz. 
Following this, there have been a number of deep searches at frequencies above 10~GHz, 
at $15$~GHz by \citet{macquart10}, at $12 - 18$~GHz by \citet{siemion13}, and 
at $19$~GHz by \citet{eatough13}.  Remarkably, despite integration times exceeding 
$10$~hours with 100-m single-dish telescopes, none of these searches has 
discovered a single pulsar in the Galactic Centre region! 

\begin{figure}[t!]
\centerline{\epsfig{file=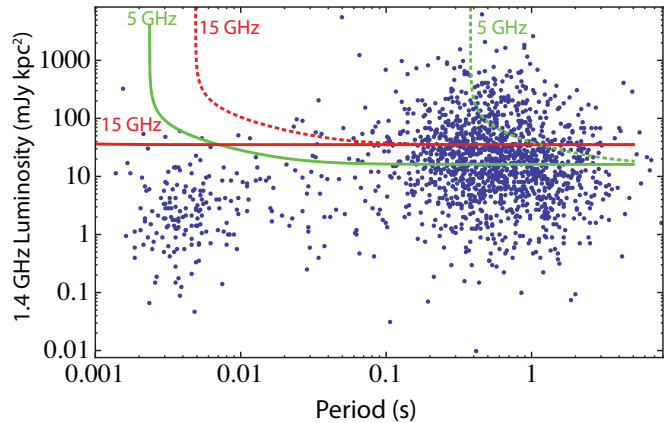,scale=0.55}}
\caption{The 1.4\,GHz luminosity (in mJy kpc$^2$) of the known pulsar population 
\citep{manchester05} is plotted versus pulsar period. The 10$\sigma$ sensitivities of 
previous 5\,GHz \citep{johnston06} and 15\,GHz GBT \citep{macquart10} searches of the Galactic 
Center are shown by the green and red curves, respectively, with the dashed and solid curves 
representing the ``strong'' and ``weak'' temporal smearing scenarios, respectively. }
\label{fig:lum}
\end{figure}

Using the temporal smearing estimates of \citet{lazio98}, \citet{macquart10} 
estimated that their 15~GHz search would have been sensitive to $\approx 15$\% of the 
Galactic center pulsar population, assuming the luminosity distribution of known 
Galactic pulsars. However, if the pulse smearing is benign, as suggested by the 
detection of the GC magnetar, then even the earlier lower-frequency searches would 
have been sensitive to normal pulsars located near the GC. Indeed, \citet{dexter14} 
estimate that both the 5~GHz search of \citet{johnston06} and the 15~GHz search 
of \citet{macquart10} would both have been sensitive to $\approx 20$\% of the GC 
pulsar population, if its characteristics resemble those of the known pulsar population 
\citep[see Fig.~2 of][]{dexter14}. This ``missing pulsar'' problem was used by 
\citet{dexter14} to argue that the GC pulsar population may be dominated by magnetars, 
i.e. that the population is very different from that in the rest of the Galaxy (where, 
as noted earlier, only $\approx 0.2$\% of known radio pulsars are magnetars, although
the magnetar birth rate may be $10-50$\% of the total neutron star birth rate in the 
Galaxy; e.g. \citealp{keane08}).

However, a number of studies have raised another possibility, namely that the GC pulsar 
population is dominated by recycled {\it millisecond pulsars} (MSPs). The dense stellar 
environment at the GC is likely to result in spinning pulsars up to millisecond periods by 
frequent close interactions with neighbouring stars, by analogy with the population of 
MSPs detected in the dense stellar environments of globular clusters 
\citep[e.g.][]{alpar82,verbunt87,camilo00,ransom05}. Note that the stellar density in the
central parsec of the GC is $\approx 10^6$ per cubic parsec \citep[e.g.][]{genzel96,schodel07}, a 
couple of orders of magnitude larger than the stellar density in globular cluster cores 
($\lesssim 10^4$~per cubic parsec), implying that close interactions are far more likely in 
the vicinity of the GC. The formation rate of low-mass X-ray binaries (LMXBs) in dense environments 
is also proportional to both the number density of neutron stars and the stellar density 
\citep{verbunt87b}. Since the neutron star density itself scales with the stellar density, 
this implies that the formation rate of LMXBs \citep[and hence, that of binary pulsars; e.g.][]{alpar82,campana98} is roughly 
proportional to the square of the stellar density. When coupled with the higher stellar density 
at the GC, this implies a far higher formation rate of LMXBs (and binary pulsars) than in 
globular cluster cores.  We note that more than $90$\% of all known globular cluster pulsars 
have periods smaller than 30~ms \citep[][]{manchester05}, indicating that a dense environment 
can dramatically alter the period distribution. Further, an over-abundance of X-ray transients 
has been detected within 1~pc of Sgr\,A* \citep[][]{muno05}; these appear to be LMXBs produced 
by three-body interactions between stellar binaries and either black holes or neutron stars 
located in the central parsec \citep[similar to the case in globular cluster cores;][]{muno05}. 
Indeed, \citet{faucher-giguere11} argue that the GC environment is likely to produce 
pulsar-black hole binary systems, via three-body interactions between stellar black holes 
and recycled pulsar binaries; the resulting pulsars would have periods in the MSP range. 
Note that, as emphasized by \citet{faucher-giguere11}, their estimate of ten times more 
MSPs in the GC region than in Terzan\,5 is likely to only be good to within an order of 
magnitude. Overall, though, it appears quite plausible that the GC pulsar population 
is dominated by recycled MSPs.

A possible caveat to the above argument is that globular clusters are not undergoing active 
star formation. As a result, neutron stars in globular clusters are ``old'' systems; 
non-recycled pulsars would have long-since spun down, so the only remaining pulsars 
are likely to be recycled MSPs. This is an important factor in the dominance of MSPs in 
the globular cluster population. In the case of the GC, there is certainly active star 
formation in the vicinity of Sgr~A*, and hence there may well be a sizeable population
of young pulsars. However, the population of neutron stars left over from the many 
earlier generations of star formation at the GC is likely to outnumber the population from 
the current generation of star formation. This earlier population is likely to have been
spun up, and would now contribute to the present MSP population. In addition, the much higher 
stellar density at the GC than in globular clusters (by about two orders of magnitude) implies 
that the scattering timescales are far shorter in the GC environment. Thus, even if there is a 
population of young pulsars in the GC environment, they would be spun up to MSPs much more 
quickly than in globular clusters; the result is that MSPs should still dominate the GC 
pulsar population.

Fig.~\ref{fig:lum} overlays the sensitivity of the best present searches \citep{johnston06,macquart10}
on the luminosity-period distribution of the known pulsar population, assuming the ``weak scattering''
scenario, i.e. that GC pulsars are subject to the same temporal smearing as the GC magnetar. While 
the 5~GHz search of \citet{johnston06} is more sensitive to normal pulsars than the 15~GHz 
search of \citet{macquart10}, even the relatively-benign assumed temporal smearing causes a 
significant reduction in its sensitivity to fast pulsars with periods $< 10$~ms. Further, 
even the 15~GHz search of \citet{macquart10} is sensitive to only a small fraction ($\lesssim 4$\%) 
of a GC MSP population whose properties resemble those of known MSPs.  We emphasize 
that this assumes the optimistic scenario in which the temporal smearing towards GC pulsars is 
similar to that measured towards the GC magnetar. The dashed curves in Fig.~\ref{fig:lum} show 
that the situation is even worse if the GC magnetar is being seen through a ``hole'' in the 
screen, and the temporal smearing towards Sgr\,A* is similar to the earlier estimates: the 
5~GHz and 15~GHz searches would then be entirely insensitive to pulsars of periods 
$\lesssim 40$~ms. Thus, given that the GC pulsar population is likely to be dominated by MSPs 
and that present searches have been insensitive to such a population, it appears that an MSP 
population is a viable way of hiding pulsars at the GC, and solving the missing pulsar problem. 


In passing, we note that the 2 brightest MSPs in Terzan\,5 would have been detected
in the 15~GHz survey of \citet{macquart10}, if located at the distance of the GC, and 
if the weak scattering scenario indeed applies to the GC pulsar population. The lack of 
detections in this survey appears surprising if the GC region indeed has $\approx 10$
times the number of MSPs seen in Terzan\,5 \citep{faucher-giguere11}. Indeed, this might 
be considered evidence that either the GC sightline is subject to strong scattering
or that the number of MSPs at the GC has been over-estimated. However, we caution 
that small number statistics makes it difficult to extrapolate from two bright MSPs 
in Terzan\,5 to the full GC population. It is hence critical to increase the sensitivity 
of GC searches so as to be able to detect a significant fraction of the Terzan\,5 
(or known MSP) population, if placed at the GC distance. If MSPs remain undetected 
in such searches, the non-detections 
would essentially imply that either the weak scattering case is not applicable to the 
GC or that the GC pulsar population is somehow very different from that in globular 
cluster environments.

\section{The detectability of pulsars at the Galactic Center} 
\label{SNSection}

In this section, we compute in detail the expected signal-to-noise ratio for MSPs at the 
Galactic Center with the most sensitive current and forthcoming radio telescopes. These computations 
are motivated both by the arguments of the preceding section concerning MSPs at the Galactic Center 
and by the dramatic increase in the fractional bandwidth of modern pulsar backends \citep[e.g.][]{siemion13}, 
due to which the frequency dependence of the relevant quantities \citep[e.g.][]{lazio98,macquart10} 
within each observing band plays a critical role in determining the optimal observing frequency. 
The following analysis takes into account variations in the signal-to-noise ratio (S/N) across 
the observing band due to the pulsar radio spectrum, the frequency dependence of the system temperature, 
and the change in the pulse width at different frequencies caused by temporal smearing.

\subsection{The S/N for wide bandwidth pulsar searches at the Galactic Center }
\label{sec:sn}

For a narrow observing bandwidth, $\Delta \nu$, the S/N for the detection of a pulsar of 
average flux density $S_\nu$, pulse period $P$ and width $W$ is \citep{lorimer06}

\begin{equation}
S/N = S_\nu \frac{G \sqrt{n_p \Delta \nu \Delta t}}{T_{\rm sys}}  \sqrt{\frac{P-W}{W}}, 
\label{SNconst}
\end{equation}
where $\Delta t$ is the telescope integration time, $n_p$ is the number of polarizations 
observed, $G$ is the telescope gain and $T_{\rm sys}$ is the total system temperature. 
Note that the above equation assumes that $S_\nu$, $T_{\rm sys}$ and $W$ do not depend on the 
observing frequency. However, when the observing bandwidth is large, this assumption breaks 
down and Equation\,(\ref{SNconst}) must be generalised to incorporate the frequency dependence 
of both the noise and the signal across the observing band.  The resulting net S/N can be estimated 
by comparing the total signal in the observing band to the total noise received during the 
duration of the observations. The total signal measured over the frequency interval $(\nu_1,\nu_2)$ in
a time period $\Delta t$ is 
then
\begin{eqnarray}
S = n_p \, \Delta t   \int_{\nu_1}^{\nu_2} S_\nu (\nu') d\nu'.
\end{eqnarray}
Next, the noise power per polarization in a single time and frequency channel of respective 
widths $\delta t$ and $\delta \nu$ is $n = T_{\rm sys}(\nu) \sqrt{\delta t \, \delta \nu} / G$.  
For observations over a number $N_{\rm ch}$ of such time, frequency and polarization channels, 
the noise adds in quadrature to give

\begin{equation}
N =   \sqrt{ \sum_i^{N_{\rm ch}} n_i^2 }.
\end{equation}

For an observation with $n_p$ polarizations, covering a frequency interval $(\nu_1,\nu_2)$, and 
of duration $\Delta t$, the above can be generalized to 

\begin{equation}
N =  \sqrt{ \Delta t \, n_p \int_{\nu_1}^{\nu_2} \frac{T_{\rm sys}^2 (\nu')}{G^2(\nu')}  \left[ \frac{W(\nu')}{P-W(\nu')}\right]  d\nu' }. 
\label{NoiseIntegral}
\end{equation}

In the above equation, the factor $W/(P-W)$ inside the integral represents the fraction of power 
present in the pulsed signal per unit time.  




Next, the telescope system temperature $T_{\rm sys}$ contains contributions from the sky and the 
telescope receiver temperature $T_{\rm rec}$, each of which has different dependences
on frequency. The sky temperature itself contains two contributions, one from the bright GC 
region, which we label $T_{\rm GC}$, and the other from the atmosphere, $T_{\rm atm}$. The 
atmospheric contribution depends on the telescope site and increases with frequency at 
frequencies $\gtrsim 20$~GHz. It is convenient to absorb the two telescope-dependent 
contributions, from the telescope electronics and the atmosphere above the site, into a 
single term which we label $T_{\rm R}$, so that the total system temperature is expressed 
in the form
\begin{eqnarray}
T_{\rm sys} (\nu) &=& T_{\rm GC}(\nu) + T_{\rm R}(\nu) 
\label{Tsys}
\end{eqnarray}

The determination of the GC contribution to the sky temperature is complicated by the fact 
that the emission near Sgr~A* has a strong spatial dependence, so that the exact amplitude 
and spectrum of the contribution depends on the telescope beam size (which itself
depends on frequency). For simplicity, we model the decline of sky temperature with 
angular distance $\theta$ from Sgr~A* using a Gaussian profile,
\begin{eqnarray}
T(\theta,\nu) = T_0(\nu) \exp \left[ - \frac{\theta^2}{\theta_0^2}\right],
\end{eqnarray}
where $T_0(\nu) = 350 (\nu/2.7\; {\rm GHz})^{-2.7}$~K \citep{reich90}. We further estimate 
$\theta_0 = 0.33^\circ$ from the 2.7~GHz image of \citet{reich90}. We emphasize that this 
approximation only applies in the neighbourhood of Sgr~A* but is sufficient for our 
purposes, given the high frequencies ($\gtrsim 5\,$GHz), and hence, the relatively small 
telescope beamwidths under consideration here. The effective contribution of the sky 
temperature to the system temperature is determined by computing the weighted average 
of the sky emission over the telescope beam, $B(\boldsymbol\theta)$:
\begin{eqnarray}
T_{\rm GC}(\nu) = \frac{\int T(\boldsymbol\theta) B(\boldsymbol\theta) d^2\boldsymbol\theta}{\int  B(\boldsymbol\theta) d^2\boldsymbol\theta } = T_0(\nu) \left( \frac{\theta_0^2}{\theta_0^2 + \theta_b^2} \right),
\end{eqnarray}
where, in the second equality, we assume a gaussian beam shape of full-width-at-half-maximum
$\theta_b = 1.22 \lambda/d$, where $d$ is the telescope diameter or, in the case of 
an interferometer, the diameter of each element of the array.
  
We have computed the S/N values for the two most sensitive present telescopes covering 
the frequency range $\approx 5-50$~GHz, the Green Bank Telescope (GBT) and the Karl G. 
Jansky Very Large Array (VLA). For these telescopes, the system temperature (not including 
the contribution from the GC region) at the low GC elevations, the telescope gain and the 
usable bandwidth in different observing bands have been obtained from the telescope webpages\footnote{\url{https://science.nrao.edu/facilities/vla} and \url{https://science.nrao.edu/facilities/gbt}}.
In addition, we also consider the proposed sensitivities of the mid-frequency array 
of the Square Kilometre Array Phase-1 (SKA1-MID) and the full Square Kilometre Array (SKA)
\citep{dewdney13}.

We consider two scattering scenarios: ``weak scattering'', where the pulse smearing timescale is assumed 
to be the same as that measured towards the GC magnetar \citep{spitler14} 
\begin{eqnarray}
\tau_{\rm smear} = 1.3 \times \left( \frac{\nu}{1 {\rm GHz}} \right)^{-4} \, {\rm s} \; ,
\end{eqnarray}
and ``strong scattering'', where the scattering occurs 130\,pc from the GC \citep{lazio98} and, using 
the GC distance and angular broadening size of Sgr A* adopted by \citet{bower14}, the pulse smearing 
timescale is given by 
\begin{eqnarray}
\tau_{\rm smear}  \approx 2.1 \times 10^2 \times \left( \frac{\nu}{1 {\rm GHz}} \right)^{-4} \, {\rm s}.
\end{eqnarray}

The observed pulse width is the sum in quadrature of the pulse scattering timescale, the intrinsic 
pulse width and the detector channel width:
\begin{eqnarray}
W = \sqrt{(w_{50} P)^2 + \tau_{\rm smear}^2 + \delta t^2}, 
\label{Weqn}
\end{eqnarray}
where the intrinsic pulse width is taken to be a fixed fraction, $w_{50}$, of the pulse 
period. The time resolution of the detector, $\delta t$, is an important factor when 
the detector time resolution is coarser than the intrinsic pulse width. In this regime, 
we see from Equation\,(\ref{Weqn}) that, in the limit of narrow pulse widths, the observed 
pulse width tends to the detector time resolution, $\delta t$, and that no further 
improvement in S/N is possible with a further decrease in $w_{50} P$. We will assume 
a time resolution of 50\,$\mu$s for the VLA, SKA1-MID and full SKA, which is more than 
sufficient to not hinder the detection of MSPs. However, in the case of the GBT, we have
assumed a time resolution of 0.5~ms, the best available with the present wideband 
VEGAS spectrometer; this does influence the detectability of the fastest MSPs, with 
periods of $\approx 1$~ms.

We note that Equation\,(\ref{SNconst}), and its generalization for large bandwidths, implies that 
a pulsar whose pulse width $W$ matches the pulse period $P$ would be completely undetectable, with 
the noise term increasing rapidly as $W$ approaches $P$. When including the effect of scatter 
broadening on the pulse width, Equation\,(\ref{Weqn}) implicitly assumes that the scatter-broadened 
pulse profile is smeared into a boxcar of width $W$, rather than being convolved with an exponential
tail. The latter case would result in some pulsed emission being visible even with $W > P$ 
\citep[e.g.][]{macquart10}. A more accurate treatment would slightly improve the detectability 
of pulsars at lower frequencies, but would not significantly alter our results.

\subsection{Application to the Galactic Center}

\begin{figure*}[htbp!]
\begin{center}
\begin{tabular}{cc}
\epsfig{file=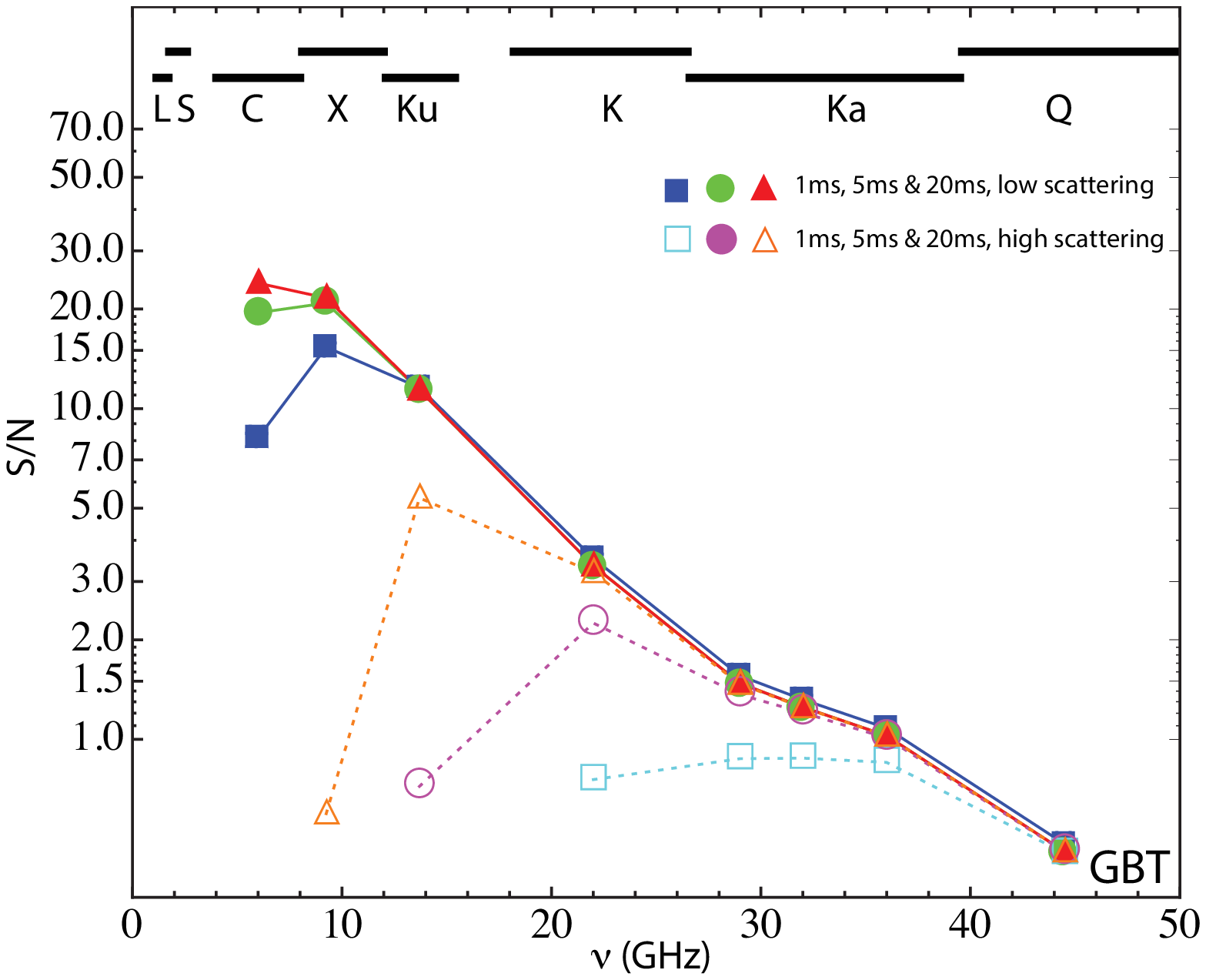,scale=0.5} &  \epsfig{file=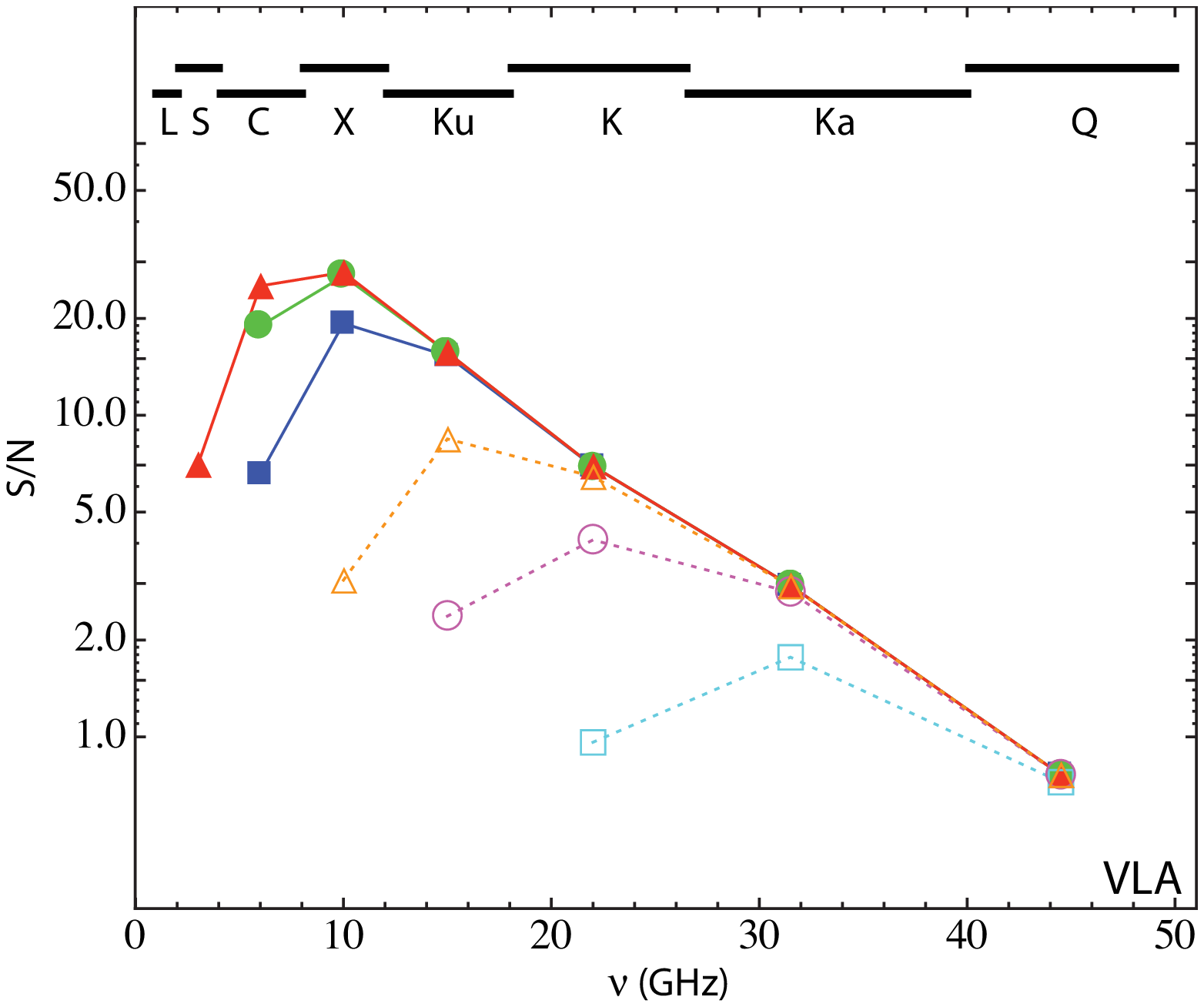,scale=0.5} \\
\epsfig{file=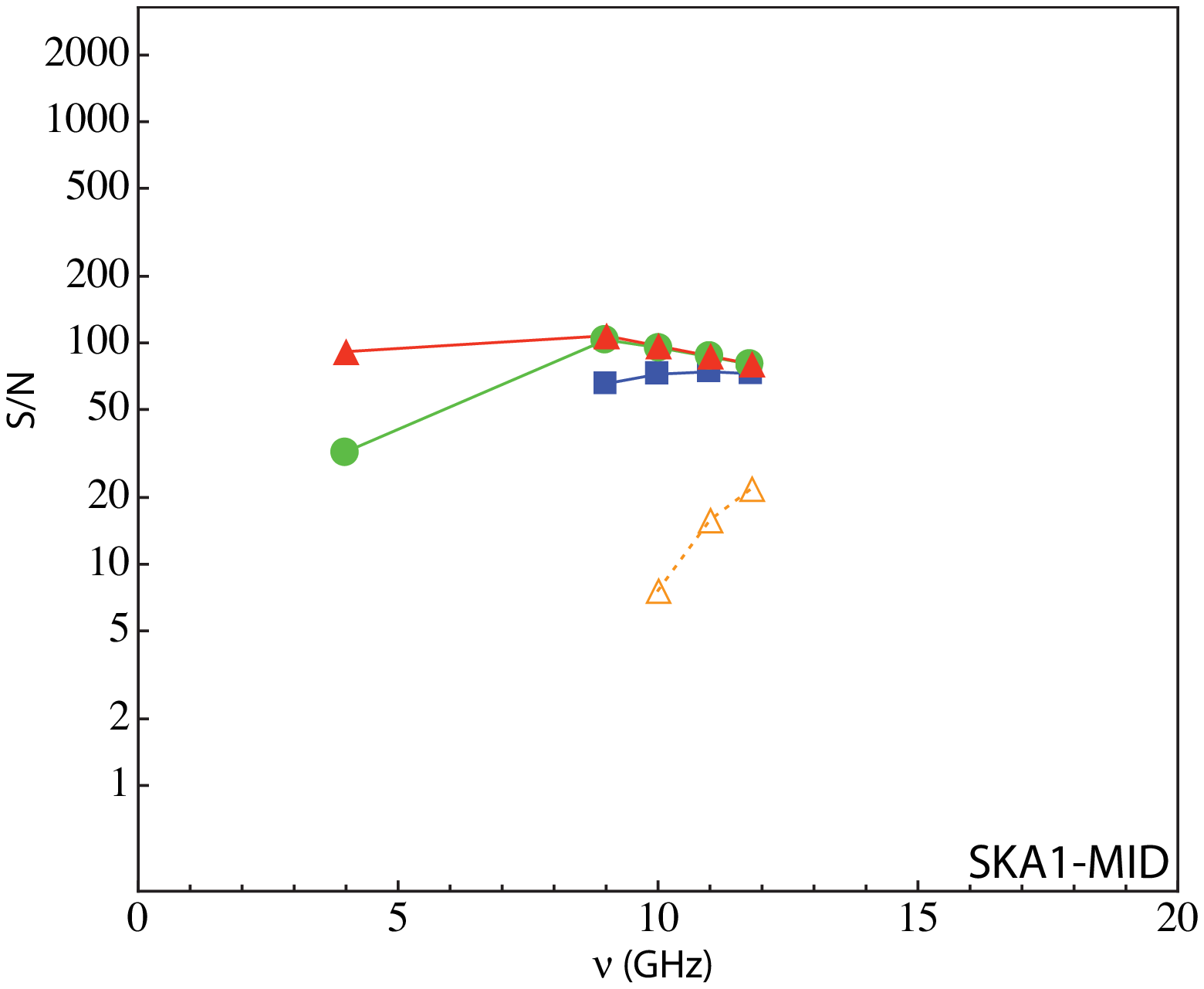,scale=0.5} & \epsfig{file=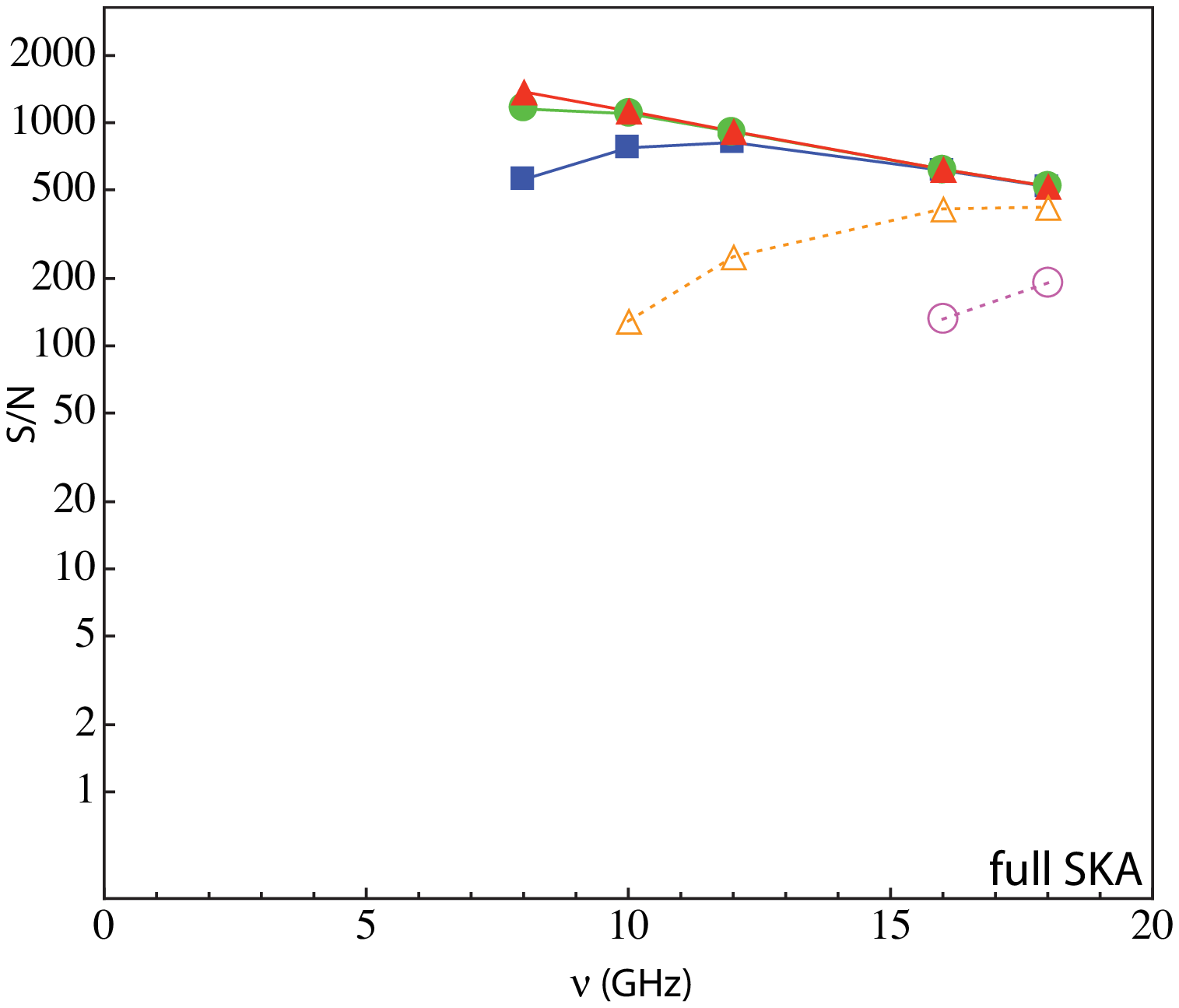,scale=0.5} \\
\end{tabular}
\end{center}
\caption{The frequency dependence of the S/N for GBT, VLA, SKA-MID and full SKA searches for MSPs 
at the Galactic Center. The filled and open symbols are, respectively, for the weak-scattering (i.e. 
magnetar-like) and strong-scattering scenarios. The symbols plotted are for MSPs with the listed periods
(1, 5, and 20\,ms), a duty cycle of 10\% and a spectral index of $-1.7$.  The vertical axis gives the 
expected S/N after a 30-hour integration for an MSP with a 1.4\,GHz luminosity of $L_{1.4} = 10\,$mJy kpc$^2$. 
The frequency ranges of the current suite of available receivers for each telescope is shown 
at the top of each panel. The optimal frequency band for MSPs with periods $\lesssim 10\,$ms is $8-12$\,GHz 
in the weak-scattering case, and $18-26$\,GHz (i.e. K-band) in the strong-scattering case.} 
\label{SensitivityFreqFig}
\end{figure*}

Incorporating all the above considerations, the signal-to-noise ratio for a pulsar at 
the Galactic Center is 

\begin{eqnarray}
{\rm S/N} = 
\frac{1.49 \times 10^{-5} \sqrt{n_p \, \Delta t}   \int_{\nu_1}^{\nu_2} L_{1.4} \,\left(  \frac{\nu }{1.4\, {\rm GHz}}\right)^{-\alpha} d\nu'}{ \sqrt{ \int_{\nu_1}^{\nu_2} \frac{T_{\rm sys}^2 (\nu')}{G^2(\nu')}  \left[ \frac{W(\nu')}{P-W(\nu')}\right]  d\nu' }},
\label{SNfulleq}
\end{eqnarray}
where we have defined the 1.4~GHz pulsar (pseudo-)luminosity $L_{1.4} = S_\nu d_{\rm GC}^2$ in units of mJy\,kpc$^2$,
the numerical prefactor assumes a GC distance of $d_{\rm GC} = 8.2\,$kpc, and $\alpha$ is the 
pulsar spectral index ($S_\nu \propto \nu^{-\alpha}$). 

Analyses of MSP spectra indicate a mean spectral index in the range $1.6$\,-\,$1.8$ \citep{kramer98,maron00}; 
we hence adopt a typical value of $\alpha = 1.7$ in our modelling. However, we note that a recent 
analysis of the slow pulsar population favours a lower spectral index, $\alpha = 1.4$ \citep{bates13}.
If this also applies to the MSP population, it would assist the detectability of pulsars at high 
frequencies. 

The optimal detection frequency for a pulsar of a given period is chiefly determined by the competition 
between scattering, which broadens the pulses to a timescale less than the pulse period only above some 
frequency $\nu_{\rm lim}$, and the pulsar spectrum, which declines with increasing frequency (of course,
secondary considerations include the sky contribution to the system temperature, the telescope 
characteristics and the observing bandwidth). It is obvious that the pulsar is undetectable at frequencies 
below which the pulse broadening timescale matches the pulsar period.  It might therefore be supposed 
that the optimal detection frequency includes frequencies just above $\nu_{\rm lim}$. However, the 
noise contribution becomes arbitrarily large for frequencies in the vicinity of $\nu_{\rm lim}$, 
where $P \approx W$.  In fact, there is a range of frequencies above 
$\nu_{\rm lim}$ where $W(\nu)$ is not sufficiently small compared to $P$, and where the contribution 
of the noise power per frequency interval in Equation\,(\ref{SNfulleq}) exceeds the contribution from the 
pulsed signal power, so that inclusion of this region decreases the S/N. It is therefore advantageous 
to restrict the lower bound of the observing band to exclude this region, so as to maximize the S/N. 
In our calculations of the S/N for bands near $\nu_{\rm lim}$, we therefore optimize the S/N by 
restricting the lower cutoff frequency to include only those frequencies which make a positive 
contribution to the S/N (i.e.\,we do not necessarily use the entire available bandwidth in the detection). 
Observationally, one would similarly attempt to maximize the S/N of any detection by restricting 
the range of frequencies only to those in which pulsed power is evident. 

The four panels of Figure~\ref{SensitivityFreqFig} illustrate the detectability of MSPs of 
three different periods, 1~ms, 5~ms and 20~ms (all with $w_{50}=10$\% and $L_{1.4}= 
10$~mJy\,kpc$^{2}$, and located at the Galactic Center distance), with a 30-hour integration 
with the four telescopes under consideration. The filled squares and solid curves correspond 
to the low-scattering magnetar-like scenario (with $\tau_{\rm smear} =1.3\nu_{\rm GHz}^{-4}$\,s), 
while the open circles and dotted curves correspond to the high-scattering scenario arising 
from a screen located 130\,pc from Sgr\,A* ($\tau_{\rm smear} = 208\,\nu_{\rm GHz}^{-4}$\,s). 
%
%

It is clear from the figure that, in the weak-scattering case, the peak S/N is obtained at a 
central frequency of $\approx 10$~GHz (i.e. the X-band for the GBT and the VLA) for MSPs with 
periods $\lesssim 10\,$ms, while for slower (20~ms) pulsars, the peak S/N shifts to 
$\approx 6$~GHz (i.e. the C-band of the GBT and the VLA), and to progressively lower 
frequencies for progressively slower rotators. Conversely, in the high-scattering case, the 
peak S/N for MSPs is obtained at much higher frequencies, $\approx 20 - 30$~GHz for 
MSPs with periods in the range $1-20$~ms.

\begin{figure}[t!]
\centerline{\epsfig{file=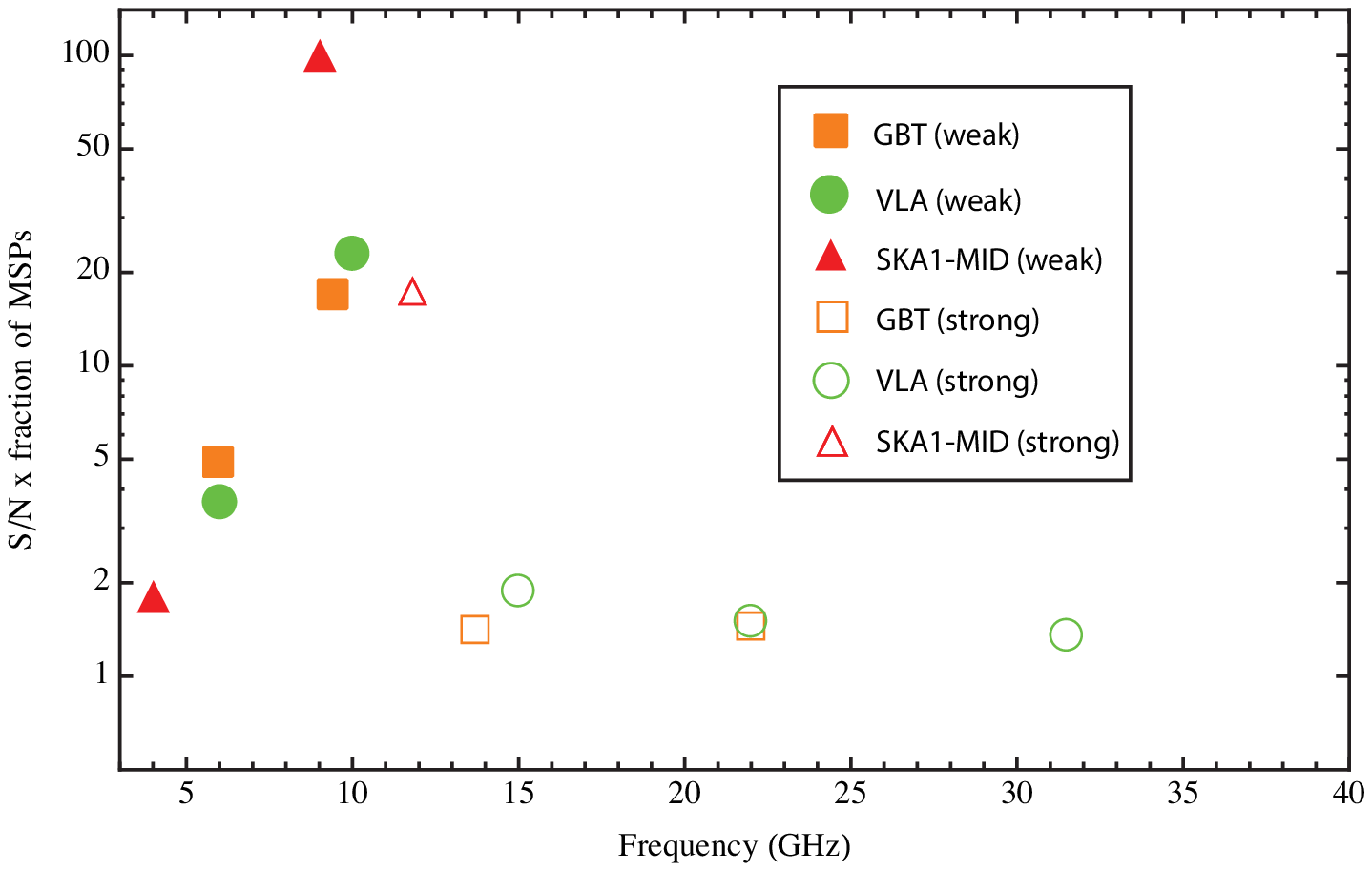,scale=0.6}}
\caption{An estimate of the optimal detection frequency for MSPs with spin periods up to 
50\,ms, under the assumption that the MSP period distribution at the GC is the same as that 
of the known MSP population. For each frequency band, the plot shows the fraction of pulsars 
with spin periods whose peak S/N falls in this band, weighted by their peak detection S/N. The 
filled symbols designate values for the weak scattering case, while open symbols indicate the 
corresponding values for the strong scattering case.}\label{fig:optimalFreq}
\end{figure}

In passing, we note that the large data volumes imply that a high statistical significance 
(typically, $\gtrsim 10\sigma$) is usually required in such pulsar surveys. This implies
that an MSP with $L_{1.4}= 10$~mJy\,kpc$^{2}$ and with periods $< 20$~ms would be detectable 
with integration times of $\approx 10-30$~hours with the GBT and the VLA (and, of course, 
with the SKA1-MID and the full SKA) in the weak-scattering case. However, much larger 
integration times ($> 100$~hours) would be needed on the GBT, the VLA or the SKA1-MID
to detect such an MSP in the high-scattering case. If the GC pulsar population is 
indeed dominated by MSPs and the high-scattering case applies, it may only be possible
to detect this population with the full SKA.

Of course, the optimal detection frequency for MSPs critically depends on both 
the MSP period distribution and the MSP luminosity function. Assuming that these
are same for the GC environment as those of the known MSP population 
\citep[e.g.][]{manchester05}, we can determine the overall optimal detection 
frequency for MSPs (integrated over all spin periods), by estimating, for each spin period, 
the frequency at which the peak S/N occurs, and then plotting this frequency against the 
peak S/N weighted by the relative fraction of MSPs with this period. The results of this analysis 
for the GBT, VLA and SKA1-MID are shown in Figure~\ref{fig:optimalFreq}, in both the weak- and
strong-scattering regimes. We note that the period distribution of the known MSP population 
is dominated by spin periods between 1 and 3\,ms. In the weak-scattering regime, 
the S/N for the three telescopes is either relatively flat over the frequency range 
$5-10$~GHz or peaks at $\approx 10$~GHz. It is hence not surprising that the optimal detection 
frequency in the weak-scattering regime closely matches that at which the S/N 
is maximal for pulsars of those periods (i.e. $\approx 8-10$~GHz). However, we note that in the 
strong-scattering regime, the optimal detection frequency is actually $\approx 15$~GHz for the VLA, 
i.e. {\it lower} than the frequency at which the S/N is maximal for MSPs with periods of $1-3$\,ms. 
This is because the VLA detection S/N in the strong-scattering regime is far higher for MSPs with 
periods of $\approx 20$\,ms than for MSPs with periods of $\approx 1-3$\,ms, and this compensates
for the larger fraction of fast MSPs in the population. 

We emphasize that the above analysis makes the critical assumption that both the MSP luminosity 
function and the MSP period distribution in the GC environment are the same as those of the known
MSP population, dominated by MSPs in globular clusters. It is not implausible that these 
are different in the GC and globular cluster environments. Interactions that form MSPs in the GC region 
are likely to differ substantially from those that operate in other known environments 
\citep[e.g. due to the increased prevalence of binary interactions in the GC region;][]{faucher-giguere11}.


Recognising that X-band is the optimal detection band for most GC MSPs in the 
weak-scattering regime, Figure~\ref{LuminosityPeriodFig} shows the limiting detection 
luminosity, scaled to 1.4\,GHz, for a 30-hour survey of the Galactic Center as a function 
of pulse period.  The 1.4\,GHz luminosities of known pulsars \citep[from the ATNF pulsar 
catalogue;][]{manchester05} are plotted for comparison. As noted earlier, previous pulsar
surveys of the GC have been insensitive to all but the most luminous members of the 
known MSP population, if located at the GC distance: the most sensitive previous search 
\citep{macquart10} would have detected only 4\% of all known pulsars with periods $\leq 20$\,ms.
By contrast, a 30\,h X-band survey with the GBT would detect 27\%, and a corresponding VLA survey 
42\%, of this population. Of course, both such surveys would, in addition, be sensitive to 
$> 80$\% of slow pulsars, with periods $\geq 0.1$\,s.


\subsection{Searches for accelerated pulsars}

Implicit in the foregoing calculations and time estimates is the assumption that it is 
possible to integrate coherently for several tens of hours on individual pulsars. This is 
potentially an issue for MSPs, which tend to occur in binary systems and where the 
inherent orbital accelerations can lead to a non-negligible pulse phase drift for 
observations lasting longer than a small fraction of the orbital period. For example, more 
than half of the known MSPs (with periods~$\leq 20$~ms) in globular clusters arise in 
binaries\footnote{See \url{http://www.naic.edu/~pfreire/GCpsr.html}}, with orbital periods 
ranging from $\approx 1.6$~hours to $\approx 191$~days \citep[e.g.][]{camilo00,ransom05},
and a median orbital period of $\approx 17$~hours. Searches for MSPs at the Galactic Center 
that extend for more than a few hours are hence likely to be affected by the above issue,
which is relevant to about half the MSP population (assuming that the distribution between ``isolated'' and 
``binary'' MSPs is the same as that in globular clusters). Pulse phase drift is a 
particularly important consideration for GC searches with the GBT and VLA, whose 
northern latitudes restrict them to individual GC observing runs of durations 
$\lesssim 6$\,hours. A 30-hour GBT or VLA integration on the GC would thus require that 
one integrate coherently over a period of at least $5$ days.  


Drifts in the pulse period over the duration of the observation must be taken into 
account, so that the sensitivity is not reduced for MSPs in short-period binaries. When 
determining the optimal method for taking the pulsar orbit into account in the 
detection process, it is convenient to characterise binary pulsars in terms of the 
ratio of total observation duration, $\Delta T$, to the orbital period, $T_{\rm orb}$. 
For pulsars with $\Delta T/T_{\rm orb} < 0.1$, the change in pulse frequency is 
sufficiently small that standard acceleration searches are efficient \citep{wood91,ransom02}. 
Conversely, if $\Delta T/T_{\rm orb} \gtrsim 1.5$, phase modulation searches may be 
employed at the penalty of a slight decrease in sensitivity \citep{jouteux02,ransom03}. 
However, in the intermediate regime $0.1 \lesssim \Delta T/T_{\rm orb} \lesssim 1.5$, 
full sensitivity searches require a search over the Keplerian parameters of the orbit. 
For orbital periods similar to those typical of binary MSPs in globular clusters, 
searches over the orbital parameters would be needed for integrations lasting longer 
than $\approx 1$~hour.

\begin{figure}[t!]
\centerline{
\epsfig{file=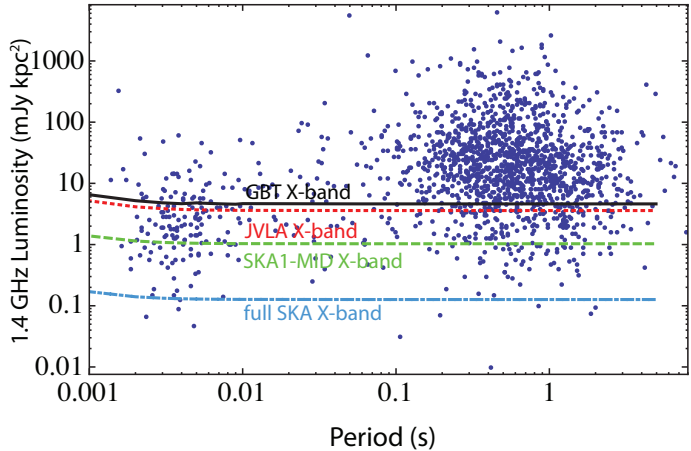,scale=0.58}}
\caption{The $10\sigma$ sensitivities of GBT, VLA, SKA1-MID and SKA 30-hour X-band integrations to the 
known Galactic pulsar population, if placed at the distance of the Galactic Center, assuming the 
weak-scattering case.  As in Fig.~\ref{fig:lum}, the dots show the 1.4\,GHz luminosity of the known 
pulsar population \citep{manchester05} plotted versus pulsar period, while  the solid, dashed, dotted
and dash-dotted curves show the $10\sigma$ sensitivities for the GBT, VLA, SKA1-MID and full SKA, 
respectively. It is clear that deep X-band observations with existing telescopes (the GBT and the 
VLA) would be sensitive to a significant fraction ($\gtrsim 30$\%) of the known MSP population (as well 
as to $\gtrsim 65$\% of the entire known pulsar population), if located at the GC distance.} 
\label{LuminosityPeriodFig}
\end{figure}

In the case of SKA1-MID and the full SKA, the relatively high sensitivity 
implies that short searches (of $\lesssim 1$~hour duration), in conjunction with 
existing acceleration search software, would be capable of detecting significant
fractions of the known MSP population ($\approx 25$\% for SKA1-MID and $\approx 90$\% 
for the full SKA) if located at the GC distance. However, for the GBT and the VLA, such 
short integrations would only be able to detect a small fraction (a few \%) of the
population, as the limiting luminosity would be a factor of $\approx 5.5$ worse 
than that of the 30-hour integrations shown in Fig.~\ref{LuminosityPeriodFig}.
One might increase the net statistical significance of detections by performing 
multiple searches over multiple short stretches of data. In such an approach, one would 
retain all candidates detected at a lower significance (i.e. $< 10\sigma$) and then 
jointly analyse all such marginal candidates to test whether the different observing 
epochs yield pulse periods and accelerations consistent with a fitted orbit.

The alternative approach would be to fit different combinations of the six Keplerian 
orbital parameters to the phase-coherent pulse data. The most effective current search 
methods in this regime have been developed and successfully implemented to detect 
binary pulsars on circular orbits \citep{knispel11,knispel13}. This technique relies 
on convolution of the data with multiple orbital parameter combinations; its extreme 
computational expense has necessitated implementation on highly distributed machines 
(e.g.\,{\it Einstein@Home}). A potential disadvantage of this approach is the 
increase in the dimensionality of the search over a conventional acceleration search, 
requiring a commensurate increase in the S/N of any detection in order to achieve an 
acceptable level of significance. However, this can be partially mitigated by the restriction 
of the search to a specific range of orbital periods and to small eccentricities, as 
has been done in previous searches.

\section{Conclusions} 
\label{ConclusionsSection}

The high stellar density in the central parsec of the Milky Way offers a solution 
to the long-standing mystery of the lack of pulsar detections at the Galactic Center.
Analogous to the situation in globular clusters (where the densities are a few orders
of magnitude lower than at the GC), the high density is likely to result in a GC pulsar 
population dominated by millisecond pulsars. Previous GC pulsar searches have been
almost entirely insensitive to such an MSP population. 

We estimate the optimal search frequency for an MSP population for two present and two future 
telescopes, the VLA, the GBT, the SKA1-Mid and the full SKA, assuming that the GC pulsar 
population has a luminosity distribution similar to that of field pulsars. We consider two 
scattering cases, weak-scattering, where the scattering screen is roughly midway between us 
and the GC, and strong-scattering, where the screen is $\approx 130$~pc from the GC. We 
find that the optimal MSP search frequencies are $\approx 8$~GHz and $\approx 25$~GHz 
in the weak-scattering and strong-scattering cases, respectively. Deep ($10-30$~hour) 
integrations with the VLA or the GBT should allow a detection of MSPs at the GC at 
$\gtrsim 10\sigma$ significance if the weak-scattering case is indeed applicable, as 
suggested by the recent detection of a magnetar close to the GC. However, the 
strong-scattering scenario would require the full SKA to detect and time MSPs at the 
distance of the Galactic Center.

\acknowledgments
Parts of this research were conducted by the Australian Research Council Centre of 
Excellence for All-sky Astrophysics (CAASTRO), through project number CE110001020. 
NK acknowledges support from the Department of Science and Technology via a Swarnajayanti 
Fellowship, and also thanks ICRAR for support during a visit during which part of this
work was carried out.  JPM thanks Yuri Levin for engaging discussions relating to the 
topic of this work. We also thank an anonymous referee for suggestions that improved 
the clarity of this paper.

\bibliographystyle{apj}
\bibliography{ms}

\end{document}